\documentclass[aps,prd,floats,amssymb,preprint,nofootinbib]{revtex4}
\usepackage{amssymb} 
\usepackage{amsmath} 
\usepackage{amstext}
\usepackage{graphicx,epsfig} 
\usepackage{epsfig} 
\usepackage{fancybox}
\usepackage{verbatim}
\usepackage[retainorgcmds]{IEEEtrantools}

\newcommand{\be}{\begin{equation}} 
\newcommand{\ee}{\end{equation}}
\newcommand{\bea}{\begin{IEEEeqnarray}{rCl}} 
\newcommand{\eea}{\end{IEEEeqnarray}}
\newcommand{\nn}{\nonumber}

\newcommand{\la}{\langle} 
\newcommand{\ra}{\rangle}
\newcommand{\half}{\frac{1}{2}}

\newcommand{\bJ}[2]{\mathrm{J}_{#1}\left( #2 \right)}
\newcommand{\bY}[2]{\mathrm{Y}_{#1}\left( #2 \right)}
\newcommand{\hd}[2]{\mathrm{H}^{(2)}_{#1}\left( #2 \right)}

\newcommand{\dta}{\mathrm{d} \tau}

\begin{document}

\title{Tackling Higher Derivative Ghosts with the Euclidean Path Integral} 
\author{Michele Fontanini$^{a,b}\!\!$ \footnote{fmichele@physics.upenn.edu}}
\author{Mark Trodden$^{a}\!\!$ \footnote{trodden@physics.upenn.edu}}

\affiliation{
$^a$Center for Particle Cosmology, Department of Physics and Astronomy, University of Pennsylvania,
Philadelphia PA 19104, USA \\
$^b$Department of Physics, Syracuse University, Syracuse NY 13244, USA}

\begin{abstract}
An alternative to the effective field theory approach to treat ghosts in higher derivative theories is to attempt to integrate them out via the Euclidean path integral formalism. It has been suggested that this method could provide a consistent framework within which we might
tolerate the ghost degrees of freedom that plague, among other theories, the higher derivative gravity models that have been proposed 
to explain cosmic acceleration. We consider the extension of this idea to treating a class of terms with order six derivatives, and find
that for a general term the Euclidean path integral approach works in the most trivial background, Minkowski. Moreover we see that even in de~Sitter background, despite some difficulties, it is possible to define a probability distribution for tensorial perturbations of the metric.
\end{abstract}

\maketitle

\section{Introduction}
The suggestion that the accelerated expansion of the universe may be explained by an infrared modification of gravity has
fueled renewed interest in higher derivative theories and their associated pathologies \cite{ArkaniHamed:2003uy,ArkaniHamed:2003uz,Piazza:2004df,Chiba:2005nz,Calcagni:2006ye,deRham:2006pe,Koyama:2007za,ArmendarizPicon:2008yv,Mannheim:2011ds}. Corrections to the Einstein--Hilbert action built with contractions of
powers of the Riemann tensor may contain four or more time
derivatives acting on the physical field, the metric. A few special cases aside, systems with more than two time
derivatives can be described via a classically equivalent
Lagrangian that quite generically contains ghosts -- degrees of freedom with the wrong-sign
kinetic terms -- leading to catastrophic instabilities if they appear in the perturbative spectrum. Unless a scheme to deal with
the ghosts is chosen, such models are therefore unsuitable to describe physical phenomena.

The best known and standard way to make sense of theories with higher derivatives is through the effective field theory approach.
From this point of view, terms that contain higher
derivatives appear from an expansion of an unknown UV--complete theory. This
expansion, by definition,
is supposed to provide an accurate description of the full theory only at
low energies, and the physical degrees of freedom are assumed to be only those that appear in the ground state of the theory~\cite{Cheung:2007st}.
Classically, the presence of extra solutions to the field
equations that correspond to the existence of ghosts
is considered as an artifact of the effective theory, due to the
truncation of an
infinite series. If then, for instance, it is possible to push the
masses of these
degrees of freedom beyond the cutoff - the energy scale below which the effective field
theory is trusted - the ghosts can be ignored.

Despite the ubiquity of the effective field theory idea, alternative procedures have been proposed to deal 
with higher derivative terms in the action~\cite{Bender:2008vh,Hawking:2001yt}.
In this paper we will focus on the prescription introduced by Hawking
and Hertog~\cite{Hawking:2001yt}, who demonstrated that the Euclidean path
integral formulation of the quantum theory allows one to define a probability distribution for a scalar field that appears in the Lagrangian with four time derivatives.

The theoretical differences between these ways of treating theories with ghosts are interesting in their own rights. However, it is important to note that if the higher order terms are considered as corrections to the second order action for the field, the results calculated in the Euclidean path integral approach could lead, in principle, to a different physical result. In fact, corrections to the probability for the fields may have a
different dependence on the ``coupling constant'' than the equivalent
corrections calculated via the effective field theory. Here by ``coupling constant'' we mean the parameter that controls the strength of the higher order term in the action. For example, in higher order theories of gravity the ``coupling constant'' contains appropriate inverse powers of the cutoff scale, and a different dependence on the behavior of such corrections may shift the energy at which they become important. This, at least in principle, holds out the hope of an observational test of these
competing ideas. However, at fourth order, the analysis in~\cite{Clunan:2009er} of a term proportional to the Weyl tensor squared
has demonstrated that in a de~Sitter background there is no discrepancy between the Euclidean path integral procedure and
effective field theory one. 

In this paper, we explore further whether the Euclidean path integral approach can be extended to apply generally, 
and to explore whether observational differences from the effective field theory approach can be realized in 
practice. Since the study of a general higher derivative correction to the Einstein-Hilbert action is prohibitively complicated, 
we therefore focus here on a nontrivial correction beyond 4th order, namely the 6th order term 
$\nabla _\alpha R _{\mu \nu} \nabla ^\alpha R ^{\mu \nu}$. As we will see, this is sufficient to draw interesting conclusions.

We demonstrate how, in principle, to apply the Euclidean path integral prescription
 to sixth order terms. However, the question of whether it can be applied to a
specific system such as General Relativity (GR) plus fourth and sixth order corrections is highly-dependent on the choice
of background. In particular, a Minkowski background always admits choices for the ``coupling constants'' that yield a
well defined Euclidean theory, while a de~Sitter background, due to its explicit time dependence, introduces some complications, since the simple requirements to apply the prescription are not met. Nevertheless, as happened in the fourth order case, we shall see that this does not preclude the possibility of finding a viable result. 

The paper is organized as follows. Section~\ref{Hawking} is devoted to a brief review of the Euclidean path integral
approach, and a discussion of its generalization to sixth order for a certain class of quadratic Lagrangians. In section~\ref{DRDR} we derive the perturbed action for tensorial modes coming from a sixth order action, about two backgrounds, Minkowski and de Sitter. We then solve for the classical 
solutions and perform the canonical procedure to build the path integral in the Lagrangian formulation.  Finally, in 
section~\ref{conclusions} we comment on the results and present our conclusions. Throughout the paper we use $t$ and $\eta$ to denote cosmological and conformal times respectively, and denote the time derivative with respect to them with an overdot $d/dt \equiv \dot{(\ )}$, with the difference between $t$ and $\eta$ being clear from the context. After a Wick rotation the time coordinate is described by a real parameter that for both cosmic and conformal times we call $\tau$, and the derivative with respect to it is represented by a prime sign $d/d\tau \equiv (\ )'$. Conformal time is only used when the de~Sitter background is taken into consideration. Greek indices run from $0$ to $3$ and Latin indices run from $1$ to $3$.

\section{Review of the Hawking-Hertog Formalism} 
\label{Hawking}

We begin by reviewing the idea behind the Euclidean path integral procedure and discussing the ways in which 
the fourth order case differs from the usual second order treatment.\\
In a second order theory the propagator for a field $\phi$ defined by a Lagrangian $L_\phi$ can be found computing a path integral between the initial and final configurations
\be
\la (\phi_f;t_f)|(\phi_i;t_i) \ra
=\int_{\phi_i} ^{\phi_f} d[\phi(t)] \exp\left[ i S[\phi] \right]  \label{} \ ,
\ee
where the action for the field is given by 
\be
S[\phi]=\int_{t_i} ^{t_f} dt' L(\dot{\phi},\phi,t') \ .
\ee
Here $\phi_\times$ represents the state of the field at time $t_\times$. A system described by a quadratic Lagrangian with a higher number of time derivatives can be transformed into a second order system via nonlinear transformations\footnote{We ignore spatial dependence for the moment or, equivalently, we think of the field $\phi$ as a particular Fourier mode.}; for instance a fourth order system with Lagrangian
\be
L=-\frac{1}{2}\phi\left(\frac{d^2}{dt^2} -m_1^2\right)\left(\frac{d^2}{dt^2} -m_2^2\right)\phi, \label{lagphi}
\ee
can be recast as 
\be
L=\frac{1}{2}\psi_1\left(\frac{d^2}{dt^2} -m_1^2\right)\psi_1-\frac{1}{2}\psi_2\left(\frac{d^2}{dt^2} -m_2^2\right)\psi_2 \ , 
\label{lagpsi}
\ee
where $\psi_1$ and $\psi_2$ are defined via
\be
\psi_1=\frac{(\frac{d^2}{dt^2} -m_2^2)\phi}{\sqrt{m_2^2-m_1^2}} \ , \ \ \ \psi_2=\frac{(\frac{d^2}{dt^2} -m_1^2)\phi}{\sqrt{m_2^2-m_1^2}} \ .
\ee
In the canonical treatment of higher order systems, this transformed Lagrangian~(\ref{lagpsi}) is the starting point and the system is viewed as a multi--field one, where at least one of the newly defined second order fields is a ghost. In the case at hand it is easy to note that $\psi_2$ has the wrong sign for the kinetic term, playing the role of the ghost field.

The propagator is given by a path integral over both fields
\be
\la (\psi_{2f},\psi_{1f};t_f)|(\psi_{2i}\psi_{1i};t_i) \ra 
= \int_{(\psi_2,\psi_1)_i} ^{(\psi_2,\psi_1)_f} d[\psi_2(t)]d[\psi_1(t)] \exp\left[ i S[\psi_2,\psi_1] \right]  \label{}\ .
\ee
Note that via the definitions of $\psi_1$ and $\psi_2$, this functional integration can be interpreted as integrating over the original field $\phi$ and its second time derivative. However, as pointed out in~\cite{Hawking:2001yt}, this choice presents a problem. For second order systems the propagator obeys the composition law
\be
G(\phi_3,\phi_1)=\int d[\phi_2] G(\phi_3,\phi_2)G(\phi_2,\phi_1) \ ,
\ee
where $G(\phi_3,\phi_1)$ is the propagator between the two states ``$1$'' and ``$3$'', and ``$2$'' represents an intermediate state. When one joins the fields above and below the intermediate time $t_2$, the value of the field $\phi(t_2)$ is fixed, but its time derivative is not, resulting in a jump in $\dot{\phi}(t_2)$, which in turn corresponds to a delta function in the value of $\ddot{\phi}(t_2)$. Unfortunately, in the original fourth order action~(\ref{lagphi}) the second time derivative appears quadratically, and hence the original composition law of the path integral is lost and infinities arise. This argument applies quite generally to higher derivative systems. In fact, the standard way to deal with fourth order systems is to use Ostrogradski's theorem~\cite{Ostrogradski:1850} to define a Hamiltonian from the fourth order Lagrangian and to take $\phi$ and $\ddot{\phi}$ to be the canonical variables over which one integrates in the path integral.

On the other hand, in the alternative procedure proposed in~\cite{Hawking:2001yt} to deal with fourth order systems, the fundamental variables are taken to be the field and its first time derivative. This choice is motivated by the need to retain the continuity properties of the path integral formulation, as described above~(see~\cite{Hawking:2001yt} for details).
However, this point of view introduces a different problem. Initial and final states are then described in terms of $\phi$ and $\dot{\phi}$, which behave much like position and momentum for a particle in quantum mechanics. The proposed procedure is to rotate the system to Euclidean time, and then to integrate out the $\dot{\phi}$ dependence in the definition of probabilities, thus obtaining well defined quantum mechanical observables at the price of a loss of unitarity. This procedure has always been possible in the special cases studied in the literature so far.

Therefore, a summary of the practical procedure is:
\begin{enumerate}
\item From the fourth order action $S$ perform a Wick rotation to obtain the Euclidean action $S^E$,
\item Derive the Euclidean equations of motion and corresponding solutions,
\item Use the Euclidean version of the path integral to find the propagator for $\phi$ with boundary conditions on $\phi$ and $\phi'$,
\item Define a ``wavefunctional'' as the propagator from a vacuum state at minus infinity in Euclidean time, 
\item Find the modulus squared of the wavefunctional, or probability amplitude, which gives the probability that a quantum fluctuation leads to a state with specified $\phi$ and $\phi'$, 
\item Finally, and crucially, trace over $\phi'$ before returning to real time. Note that if one were to rotate back to Lorentzian time before integrating, the probability would be ill defined, reflecting the existence of the ghost degree of freedom.
\end{enumerate}

There is no magic trick behind all this, since taming the ghost by integrating over the infinities that it introduces happens at the price of a violation of unitarity\footnote{See the original paper~\cite{Hawking:2001yt} for a detailed discussion.}. The Euclidean formulation of the path integral together with the requirement that the fields die off at Euclidean infinity ensures that the fields remain bounded in real time. This is
similar to using a final boundary condition to remove runaway solutions from systems that would otherwise contain them.

Let us examine this procedure in the specific case of the 4th order system discussed earlier. Using $t$ for Lorentzian and $\tau$ for Euclidean time, rescaling the field $\phi$, the action is
\be
S=\int dt \left( \frac{\alpha^2}{2} \ddot{\phi}^2-\frac{1}{2}\dot{\phi}^2+\frac{m^2}{2}\phi^2 \right) \ ,
\ee
where $\alpha^2/2$ is an arbitrary small parameter, the ``coupling constant'' mentioned earlier. After a Wick rotation $t\rightarrow i \tau$ the action becomes
\be
S^E\equiv \int d\tau \left( \frac{\alpha^2}{2} \phi''^2+\frac{1}{2}\phi'^2+\frac{m^2}{2}\phi^2 \right) \ ,
\ee
so that $iS \rightarrow - S^E$. When $S^E$ is positive definite the path integral converges giving a well defined Euclidean quantum theory. The resulting equations of motion take the form $D_4 \phi = 0$, where
\be
D_4 = \frac{1}{2} \left(\alpha^2 \frac{\mathrm{d}^4}{\dta ^4} -
\frac{\mathrm{d}^2}{\dta ^2} + m^2 \right) \ ,
\ee
and admit solutions 
\be
\phi(\tau) = A_1 \sinh(\lambda_1 \tau)+ A_2 \cosh(\lambda_1 \tau)
+ A_3 \sinh(\lambda_2 \tau)+ A_4 \cosh(\lambda_2 \tau) \ . 
\label{eomsol}
\ee
The path integral for the propagator from state $(\phi_1,\phi_1')$ at Euclidean time $-T$ to the state $(\phi_2,\phi_2')$ at Euclidean time $0$ is then
\bea 
\la (\phi_0,\phi_0';0)|(\phi_T,\phi_T';-T) \ra
&=&\int_{(\phi_T,\phi_T')} ^{(\phi_0,\phi_0')} d[\phi(\tau)] \exp\left[
-S^E[\phi] \right] \nn \\
&=& e^{ -S^E[\phi_{\rm cl}]} \int_{(0,0)} ^{(0,0)}
d[\varphi(\tau)] \exp\left[ -S^E[\varphi] \right] \ ,
\label{pidef4}
\eea 
where we have used the decomposition $\phi=\phi_{\rm cl} + \varphi$, with
$\phi_{cl}$ the classical solution of the Euclidean equations of motion for the appropriate boundary conditions.

The wavefunctional for a state described by $(\phi_0,\phi_0')$ at time $\tau=0$ is then defined via~(\ref{pidef4}) as
\be
\Psi_0[\phi_0,\phi_0'] \equiv \lim_{T\rightarrow \infty} \la (\phi_0,\phi_0';0)|(\phi_T,\phi_T';-T) \ra \ ,
\ee
which yields
\be
\Psi_0[\phi_0,\phi_0']=N \exp \left[ -A \phi'^2 _0+B \phi_0 \phi_0' -C\phi_0^2 \right] \ .
\ee
The values of the coefficients $A$, $B$, and $C$, can be found in the appendix, and $N$ is a normalization factor found by calculating the path integral over the field $\varphi$. There has been some debate in the literature about how to actually calculate this normalization function, and we refer the interested reader to the very clear article by Zerbini and Di Criscienzo~\cite{DiCriscienzo:2009pb}, and references therein, for a complete discussion.

The next step is to define a probability 
\be
\bar{P}[\phi_0,\phi_0'] \equiv \Psi_0 \Psi^* _0= N^2 \exp\left[-2 A\left( \phi'^2_0+\frac{m}{\alpha} \phi_0^2\right) \right] \ .
\ee
As we have already mentioned, this would not provide a well defined probability if rotated back to Lorentzian time, since $A>0$ and the rotation would introduce $(-i)^2=-1$ in front of the $\phi_0'$ term. Therefore one rotates back to real time only after integrating over $\phi'$, to yield as $\alpha \to 0$ 
\be
P[\phi_0]=\sqrt{\frac{m}{\pi}\left(1+ m \alpha +\dots \right) } \exp\left[ -m\left( 1+m \alpha +\dots \right) \phi_0^2 \right] \ ,
\ee
after normalizing the probability density.

How might this procedure be extended to an arbitrary higher order system with a quadratic Lagrangian? Since ultimately we wish to consider higher order terms as corrections to the propagation of the degrees of freedom of a second order Lagrangian, we seek a way to generalize this procedure so that an integration over all the extra degrees of freedom is performed in order to obtain the final results. Although some of the original motivations presented in~\cite{Hawking:2001yt} for taking fourth order terms seriously are lost in this approach, this point of view is nonetheless consistent with the proposed procedure since it corresponds to tracing over the unobserved degrees of freedom. Guided by the need for a composition law for the path integral, we are led to consider the metric perturbation $\gamma_{ij}$ and its derivatives $\gamma'_{ij}$, and $\gamma''_{ij}$ (rather than $\gamma$, $\gamma''$ and $\gamma^{VI}$) as the dynamical degrees of freedom in a sixth order Lagrangian for Gravity. The rest of the procedure developed in~\cite{Hawking:2001yt} is then unmodified, and in principle the only difficulties that appear should be those associated with the explicit calculation of the normalization function for the wavefunctional.


\section{Sixth Order Corrections} 
\label{DRDR}

Since fourth order corrections have already been analyzed in~\cite{Clunan:2009er}, we focus here on calculating
the corrections to the tensor part of the two point function coming from a sixth order term.

\subsection{Expanding the action} 

Our goal is to take a convenient contraction of Riemann tensors and their derivatives,
and to expand it to quadratic order in perturbations about a conformally flat background. We will then study the action for the perturbations around
two important backgrounds -- Minkowski space and de~Sitter space.

We focus on one of the simplest covariant terms that contains six time derivatives and is quadratic in metric perturbations,
\be
\nabla _\alpha R _{\mu \nu} \nabla^\alpha R ^{\mu \nu} \ .
\ee
The total action we start from therefore consists of the Einstein-Hilbert term, a cosmological constant, two distinct fourth order contributions and the term above
\be
S=\int d^4 x \sqrt{-g} \left[ M_{pl}^2  \left(\frac{R}{2}+\Lambda \right) +\lambda^2 R^2  -\alpha^2 C_{\mu \nu \rho \sigma}C^{\mu \nu \rho \sigma}
-  \frac{\beta^2}{M_{pl}^2} \nabla _\alpha R _{\mu\nu} \nabla ^\alpha R ^{\mu \nu} \right] \ ,
\label{fullaction}
\ee
where $\Lambda=0$ for a Minkowski background, and is nonzero for a de~Sitter one. While this action is quite general, we shall henceforth ignore the $R^2$ term; its presence does not affect the result as we have explicitly checked, and as one would expect since it merely corresponds to an additional massive scalar degree of freedom. This can be seen by changing frame via a conformal transformation of the metric.

Writing the flat Friedmann, Robertson-Walker (FRW) metric in terms of conformal time $\eta$, the perturbed metric is then
\bea 
g_{\mu \nu}&=& g^0_{\mu \nu} + h_{\mu \nu} \nn \\ 
&=&e ^{2\rho} \left( \eta _{\mu \nu} + \delta^i _\mu \delta^j _\nu
\gamma _{i j}\right) \ ,
\eea
where the scale factor $e^{\rho(\eta)}$ is equal to one for Minkowski space and equal to $(-H\eta)^{-1}$ in de~Sitter space. Since the perturbation $\gamma _{i j}$ is traceless and divergenceless $\gamma_{ii}=\partial _i \gamma _{i j}=0$,  the first non zero term in the perturbed action is\footnote{We discuss this expansion in the appendix.} 
\bea
S_\gamma &=& \int d\eta d^3x \Bigg[  -\frac{\beta^2}{2M_{pl}^2} e^{-2\rho}  \big[ -(\gamma^{III})^2+\ddot{\gamma}^2 (-10\dot{\rho}^2+4\ddot{\rho})
+3 \ddot{\gamma}^2_{,i} -3\dot{\gamma}^2_{,ij}+2\dot{\gamma}^2_{,i}(\ddot{\rho}+2\dot{\rho}^2) \nn \\
&+&\dot{\gamma}^2(4\rho^{IV}-4\dot{\rho}\rho^{III}-20\ddot{\rho}^2
+44\dot{\rho}^2\ddot{\rho}-48\dot{\rho}^4)
+\gamma_{,ijl}^2 +6\gamma_{,ij}^2(\dot{\rho}^2-\ddot{\rho}) \nn \\ &+&\gamma_{,i}^2 (-4\rho^{IV} -24\dot{\rho}^4 -8\ddot{\rho}^2 +72\dot{\rho}^2 \ddot{\rho} -4\dot{\rho} \rho^{III} ) \nn \\
 &+&\gamma^2(-8\rho^{VI}+336\dot{\rho}^4\ddot{\rho}+64\rho^{IV}\ddot{\rho}
+56\ddot{\rho}^3
-48\dot{\rho}^2\rho^{IV}-688\dot{\rho}^2\ddot{\rho}^2-304\dot{\rho}^3 \rho^{III}
+36(\rho^{III})^2 \nn \\
&&+56\dot{\rho}\rho^{V}+104\dot{\rho}\ddot{\rho}\rho^{III}+72\dot{\rho}^6)\big]
- \alpha^2 \left( \ddot{\gamma}^2-2\dot{\gamma}^2_{,i}+\gamma^2_{,ij} \right) \nn \\
&+&\lambda^2 \big[ 6\dot{\gamma}^2 (\dot{\rho}^2+\ddot{\rho})
-6\gamma^2_{,i}(\dot{\rho}^2+\ddot{\rho})
+\gamma^2(-12\rho^{IV}+72\dot{\rho}^2\ddot{\rho}
+12\dot{\rho}\rho^{III}-6\ddot{\rho}^2-18\dot{\rho}^4)\big] \nn \\
&+& \left. \frac{M_{pl}^2}{2}e^{2\rho} \big[ \dot{\gamma}^2-\gamma^2_{,i}-\gamma^2(\dot{\rho}^2+2\ddot{\rho}) \big] \right] \ . 
\label{action6} 
\eea
Note that the background equations have not been used in this derivation. We now specialize to the two cases of interest.

\subsection{Minkowski background}
\label{Minkowski_bg}

Performing a Wick rotation to imaginary time, and focusing on a Minkowski background, for which $e^{\rho(\eta)}=1$, the full sixth order action~(\ref{action6}) 
reduces to 
\bea
S^E_M=-\int d\tau d^3x \left[ -\frac{\beta^2}{2M_{pl}^2} \left(\gamma'''^2+3\gamma''^2_{,i} +3\gamma'^2_{,ij} +6\gamma_{,ijl}^2 \right) \right.
&&- \alpha^2 \left( \gamma''^2 +2\gamma'^2_{,i}+\gamma^2_{,ij} \right) \nonumber \\
&&- \left. \frac{M_{pl}^2}{2} \left(\gamma'^2 +\gamma^2_{,i} \right) \right] \ ,
\eea
where, for simplicity, we have omitted the indices on, and the argument of the perturbation $\gamma_{ij}(\eta)$. It is convenient to treat the problem in 
momentum-space by performing a Fourier transform on $\gamma$
\be
\gamma_{ij}(\eta,\vec{x})=\int \frac{d^3k}{(2\pi)^3} \sum_{s=\pm}\epsilon_{ij}^s(\vec{k})\gamma^s_{\vec{k}}(\eta) e^{i\vec{k} \cdot \vec{x}} \ ,
\ee
where the polarization tensor satisfies $\epsilon_{ii}=k^i\epsilon_{ij}=0$, $\epsilon^*_{ij}(\vec{k})=\epsilon_{ij}(-\vec{k})$, and $\epsilon^s_{ij}(\vec{k}) \epsilon^{r*}_{ij}(\vec{k})=2 \delta^{sr}$.

In order to avoid confusion through notation, we will drop all the unnecessary indices. The action for the $k$--mode then becomes
\bea
S^E_{M \, k}=\int d\tau \left[  -\frac{\beta^2}{M_{pl}^2} \left(-|\gamma'''|^2+3k^2 |\gamma''|^2
-3k^4 |\gamma'|^2 +6 k^6 |\gamma|^2 \right) 
\right. && \left. - 2 \alpha^2 \left( |\gamma''|^2-2k^2 |\gamma'|^2+|\gamma|^2 \right) \right.  \nonumber \\
&&\left. +  M_{pl}^2 \left(|\gamma'|^2-k^2 |\gamma|^2 \right)  \right] \ ,
\eea
where we have used the notation $|\gamma^{(n)}|^2\equiv \frac{d^n}{d\tau^n} \gamma \frac{d^n}{d\tau^n} \gamma^*$. Varying this action with respect to 
$\gamma^*$ yields the Euclidean equations of motion
\be
D^M_6 \gamma(\eta)=0 \ ,
\ee
with
\bea
D^M_6\equiv\frac{d^6}{d\tau^6}-\left(3k^2+\frac{2\alpha^2 M_{pl}^2}{\beta^2} \right)\frac{d^4}{d\tau^4} &&+\left( 3 k^4+4k^2\frac{\alpha^2 M_{pl}^2}{\beta^2}+\frac{M_{pl}^4}{\beta^2}\right)\frac{d^2}{d\tau^2} \nn \\
&&-\left(  k^6+2k^4 \frac{\alpha^2 M_{pl}^2}{\beta^2}+k^2\frac{M_{pl}^4}{\beta^2} \right)
\eea
Solutions to these equations can easily be written in terms of exponentials as
\be
\gamma_{cl}^M (\tau)= c_{11} e^{\lambda_1 \tau}+ c_{12} e^{-\lambda_1 \tau}
+c_{21} e^{\lambda_2 \tau}+c_{22} e^{-\lambda_2 \tau}+c_{31} e^{\lambda_3 \tau}+c_{32} e^{-\lambda_3 \tau} \ ,
\ee
with $\lambda_1$, $\lambda_2$, and $\lambda_3$ given by 
\bea
\lambda_1 &=& k \ , \nonumber \\
\lambda_{2,3} &=& \sqrt{k^2+\frac{M^2_{pl} \alpha ^2}{\beta ^2}\pm \sqrt{\frac{M^4_{pl} \left(\alpha ^4-\beta ^2\right)}{\beta ^4}}} \ .
\eea

Following the procedure highlighted in the previous section we now define a wavefunctional that describes the probability amplitude of 
being in a state characterized by $\gamma_0$, $\gamma'_0$ and $\gamma''_0$
\bea
\Psi^E_{0 M}[\gamma_0,\gamma'_0,\gamma''_0]&=&N e^{-S^E_M[\gamma_{cl}]} \nn \\
&=&N  \exp \Bigg[- \frac{1}{2M_{pl}^2} \left( A_{00} \gamma^*_0 \gamma_0  \right. + A_{01} \gamma^*_0 \gamma'_0 + A_{02} \gamma^*_0 \gamma''_0  
+A_{10} \gamma'^*_0 \gamma_0 \nn \\
&&+ A_{11} \gamma'^*_0 \gamma'_0 
+ \left.  A_{12} \gamma'^*_0 \gamma''_0 + A_{20} \gamma''^*_0 \gamma_0 +  A_{21} \gamma''^*_0 \gamma'_0+ A_{22} \gamma''^*_0 \gamma''_0 \right) \Bigg] \ .
\label{PsiM6}
\eea
The coefficients $A_{jl}$ are functions of the three $\lambda_i$, and we present their explicit forms in the appendix. It is, in fact, possible to calculate the normalization factor $N$ using Forman's theorem~\cite{Forman:1987}.
However, since this does not change our result, for simplicity we shall ignore the contributions coming from $N$ in what follows, until a normalization for the probability is needed.

A probability distribution for $\gamma_0$ can then be defined integrating over $\gamma''_0$ and $\gamma'_0$ and by rotating back to Lorentzian time
\be
P^E [\gamma_0] \equiv \int d[\gamma'_0] \int d[\gamma''_0] \Psi^E_{0 M} \Psi^{E *}_{0 M}  \rightarrow P[\gamma_0] \ ,
\ee
where the arrow implies rotating clockwise in the complex plane to Lorentzian time. The normalized probability expanded for $M_{pl} \gg 1$ then gives
\bea
P[\gamma_0]=\left( M_{pl}\sqrt{\frac{k}{\pi}}+\dots \right) \exp \Bigg[ - k M_{pl}^2 &\Bigg(&1+\frac{k \left(2 \alpha ^2+\beta \right)}{M_{pl} \left(\sqrt{\alpha ^2-\sqrt{\alpha ^4-\beta ^2}}+\sqrt{\alpha ^2+\sqrt{\alpha ^4-\beta ^2}}\right)} \nn \\
&&+\dots \Bigg) |\gamma_0|^2 \Bigg] \ .
\eea
Interestingly, we have encountered no difficulties in extending the Euclidean path integral prescription to our sixth order term in a
Minkowski background. This straightforward extension suggests that it may be possible to extend the procedure to any system with $2n$ derivatives. 

\subsection{The de~Sitter background}

We now repeat the above calculation in a de~Sitter background. As we shall see, the explicit time-dependence of the background introduces crucial differences in this case. Setting $\Lambda > 0$ and the scale factor to be $e^\rho=(-H\eta)^{-1}$, the action in Euclidean time and Fourier space becomes
\bea
S^E_{dS \ k}=\int d\tau \Bigg[ &&\beta^2 \frac{H^2}{2M_{pl}^2}\tau^2 \left[ \gamma'''^2 +\gamma''^2 \left( 3k^4+6\frac{k^2}{\tau^2}+\frac{8}{\tau^4} \right)+\gamma^2\left(k^6+8\frac{k^2}{\tau^4}\right) \right]  \nn \\
&&- \alpha^2 \left( \gamma''^2 +2k^2 \gamma'^2+k^4 \gamma^2 \right) + \frac{M_{pl}^2}{4H^2 \tau^2} \left(\gamma'^2 +k^2 \gamma^2 \right) \Bigg] \ .
\eea
Note that if we started without the sixth order term (i.e. set $\beta=0$) we would have the action presented in~\cite{Clunan:2009er}, which is not positive definite. 
Nevertheless, the authors of~\cite{Clunan:2009er} showed that this does not prevent one from following the Euclidean path integral procedure and obtaining a well defined result. We will therefore adopt the same point of view here and, although we realize that we are dealing with a non positive definite Euclidean action, proceed as planned to see if a meaningful result can be obtained.

It can also be noted that in principle we could obtain a positive definite action if we started from a different form for equation~(\ref{fullaction}). There, in fact, the signs of $\alpha^2$ and $\beta^2$ have been chosen arbitrarily. If we were to change the signs though, the results presented in section~\ref{Minkowski_bg} would not stand. We choose to keep the sign conventions so that the validity of the method is preserved in a Minkowski background.

Defining, for simplicity, $z=-k\tau$, the Euclidean equations of motion become
\be
D_6^{dS} \gamma(z)=0 \ ,
\label{dSEOM}
\ee
with
\bea
D_6^{dS}\equiv\frac{d^6}{dz^6}&+&\frac{6}{z}\frac{d^5}{dz^5}+\left(-3+\frac{C_1}{z^2} \right)\frac{d^4}{dz^4}-\frac{12}{z}\frac{d^3}{dz^3} 
+\left( 3+\frac{(4-2C_1)}{z^2}+\frac{C_2}{z^4}\right)\frac{d^2}{dz^2} \nn \\
&+&\left( \frac{6}{z}-2\frac{C_2}{z^5}\right) \frac{d}{dz}-\left(1-\frac{C_1}{z^2}+\frac{C_2}{z^4} \right) \ ,
\label{dSD6}
\eea
where 
\bea
C_1&=&2\left(\frac{\alpha M_{pl}}{\beta H}\right)^2 \\
C_2&=&8+24\left(\frac{\lambda M_{pl}}{\beta H}\right)^2+\frac{1}{2}\left(\frac{M_{pl}^2}{\beta H^2} \right)^2 \ .
\eea
Solutions to these equations can be found by factorizing the sixth order differential operator\footnote{For details see the appendix.} $D_6^{dS}$, and can be written in terms of exponentials and Bessel functions as
\bea
\gamma_{cl}(z)=&&A_1\left[ \sinh{(z)} - z \cosh{(z)} \right] +A_2\left[z \sinh{(z)} -\cosh{(z)} \right]\nn \\
&&+A_3 z^{\frac{3}{2}} \bJ{\lambda_1}{-i z}+A_4 z^{\frac{3}{2}} \bY{\lambda_1}{-i z} 
+A_5 z^{\frac{3}{2}} \bJ{\lambda_2}{-i z}+A_6 z^{\frac{3}{2}} \bY{\lambda_2}{-i z} \ ,
\label{dSsol}
\eea
where $\mathrm{J}$ and $\mathrm{Y}$ are respectively Bessel functions of first and second kind. Recalling that $z$ takes values in $(0,+\infty )$ with $+\infty$ being the past infinity boundary, in order to find the wavefunctional we need to apply a set of boundary conditions analogous to the one described earlier, namely
\be
\left.
\begin{array}{r}
\gamma(z) \rightarrow 0 \\
\gamma'(z) \rightarrow 0 \\ 
\gamma''(z) \rightarrow 0
\end{array}
\right\} \quad z\rightarrow +\infty \quad \mathrm{and} \quad \left.
\begin{array}{r}
\gamma(z) \rightarrow \gamma_0 \\
\gamma'(z) \rightarrow \gamma'_0 \\ 
\gamma''(z) \rightarrow \gamma''_0 \\
\end{array}
\right\} \quad z\rightarrow +z_0 \ .
\ee
The relevant classical solution of the equations of motion is therefore
\be
\gamma_{cl}(z) = B_1\left(1+z\right)e^{-z} +B_2 z^{\frac{3}{2}} \hd{\lambda_1}{-i z}
+B_3 z^{\frac{3}{2}} \hd{\lambda_2}{-i z} \ ,
\ee
where $\mathrm{H}^{(2)}$ represents the Hankel function of the second kind, and the coefficients $B_i$ contain the dependence on $z_0$ and on the boundary conditions $\gamma_0$, $\gamma'_0$, and $\gamma''_0$.

To calculate the wavefunction it is sufficient to rewrite the action as
\be
S_{dS k}^E=\left[\mathrm{surface\, terms} \right]+\int_{-\infty}^{\tau_0} d\tau \gamma D_6^{dS} \gamma \ ,
\ee
so that on the classical path only the first set of terms survives, with the contribution from the integral term being zero. Since we are ultimately interested in integrating over $\gamma''_0$ and $\gamma'_0$ it is convenient to collect terms and write the wavefunctional schematically as
\bea
\Psi_0^{dS}=N \exp \Bigg[ -i \frac{k^3}{D} \left( A_{00} \gamma^*_0 \gamma_0  \right. &+& A_{01} \gamma^*_0 \gamma'_0 
+ A_{02} \gamma^*_0 \gamma''_0  +A_{10} \gamma'^*_0 \gamma_0+ A_{11} \gamma'^*_0 \gamma'_0 \nn \\
&+& \left.  A_{12} \gamma'^*_0 \gamma''_0 + A_{20} \gamma''^*_0 \gamma_0 +  A_{21} \gamma''^*_0 \gamma'_0+ A_{22} \gamma''^*_0 \gamma''_0 \right) \Bigg] \ .
\label{PsidS}
\eea
The analytic dependence  of the coefficients $A_{ij}$ and $D$ on the parameters $\alpha$, $\beta$, and $M_{pl}/H$ appearing in the action is somewhat complicated and not very instructive, and so we do not display this here. 

To make progress analytically we now introduce an approximation scheme, taking $\alpha$, $\lambda$ (if the $R^2$ term is considered) and $\beta$ to be of order unity, with $M_{pl}/H \ll 1$ playing the role of the small parameter in a series expansion. Beside the reasonable choices for the parameters in the action, an extra assumption is needed to simplify the calculation. We assume that $\beta^2 < 2 \alpha^4$, allowing us to approximate the frequencies $\lambda_1$ and $\lambda_2$ and the Hankel functions. With these approximations the associated probability takes a form similar to that of equation~(\ref{PsidS}), with the same kinds of terms and different coefficients. In particular, focusing on the coefficient of $\gamma''^*_0 \gamma''_0$, which we require to have a negative real part in order to proceed with the integration, we find
\be
\bar{\bar{P}}[\gamma_0,\gamma'_0,\gamma''_0] \equiv N N^* \exp \left[ -\frac{\alpha ^2 k^3  \tau ^4 \left( 1+\sqrt{1-\frac{\beta^2}{2\alpha^4}}\right)}{k^2 \tau ^2-1} \gamma''^*_0 \gamma''_0 + \dots \right] \ .
\label{dSPbb}
\ee
$\bar{\bar{P}}$ is not yet the probability we are looking for, since integration over $\gamma''_0$ and $\gamma'_0$ is still needed. The bars are a reminder of this fact, counting the maximum number of derivatives acting on $\gamma_0$. From equation~(\ref{dSPbb}) we note that gaussian integration over the real and imaginary parts of $\gamma''_0$ can be performed only if $(k\tau)^2 >1$. Recalling that $k^2\eta^2=k^2/(a H)^2$, with $a$ being the scale factor, considering $k^2\tau^2>1$ means that the treatment can be considered valid for subhorizon modes.

With the above assumptions both the integrations over $\gamma''_0$ and $\gamma'_0$ can be performed, and after rotating back to Lorentzian time the full final result is reported in the appendix. Before we can say we have found a probability for $\gamma_0$, one last check is necessary: the coefficient of $|\gamma_0|^2$, in Lorentzian time, has to be negative in order to have a well defined (normalizable) probability. We check this by expanding the argument of the exponential as a series in $M_{pl}/H$, keeping only the leading contribution
\be
P_{L}[\gamma_0]=\tilde{N} \tilde{N}^* \exp \left[ \left(-\frac{k^3 \left(1+2   \sqrt{\frac{2\alpha ^4}{\beta ^2 }}\right)}{2  \left(1+k^2 \eta ^2\right)}
\frac{M_{pl}^2}{H^2}
 + O\left(\frac{M_{pl}}{H}\right) \right) \gamma^*_0 \gamma_0\right] \ ,
\ee
where the symbol $L$ is a reminder that we have rotated back to Lorentzian time. We can see that the probability can be integrated over all values of $|\gamma_0|$ giving a sensible extension of the method in~\cite{Hawking:2001yt} to the sixth order case. This may be compared with the equivalent form for the probability in GR,
\be
P_{GR}[\gamma_0]=|\hat{N}|^2 \exp \left[ -\frac{k^3 M_{pl}^2}{2H^2\left(1+k^2 \eta^2 \right)} |\gamma_0|^2 \right]
\ee

Finally, from the probability distribution we obtain the two point function for the tensorial perturbations $\gamma_0$ in the sixth order case
\be
\la |\gamma_0|^2 \ra \simeq \left( \frac{H}{M_{pl}}\right)^2\frac{1+k^2 \eta^2}{k^3\left( 1+ 2\sqrt{\frac{2\alpha^4}{\beta^2}}\right)} \ .
\ee

\section{Conclusions}
\label{conclusions}

The Euclidean path integral prescription is a method to integrate out the infinities appearing in higher derivative theories with ghosts and extract meaningful probability distributions for the non-ghost degrees of freedom. In this paper we have reviewed the original fourth order version of the method and have shown how to extend this to a sixth order system in a Minkowski background and in a time dependent one -- de~Sitter. The two cases are treated separately since we have shown that a time dependent background, even if highly symmetric, introduces some difficulties. The Euclidean action is in fact not positive definite, raising doubts about the validity of the underlying quantum theory. Fortunately, as in the fourth order case, this does not prevent us from finding a sensible result.

With higher order gravity in mind, in this paper we have examined an action containing GR, a sixth order term and two fourth order ones, with relative strengths set by the Planck mass and their relative mass dimension. We have found that the Euclidean path integral prescription can be applied to find corrections to the probability distribution of the tensorial perturbations about both Minkowski and de~Sitter backgrounds. The corrections we have found are at least of order one in the de~Sitter case, depending on the values of the parameters appearing in the action. Therefore the results pose stringent constraints on either the validity of the approach, or the presence of the covariant sixth order term considered.

It is important to be clear about the assumptions made throughout this paper. The first one has already been mentioned, and concerns the validity of the quantum theory when the Euclidean action is not positive definite. However, note that we could have performed the whole calculation in Lorentzian signature, and the present procedure is merely an ad~hoc prescription for rotating to Euclidean signature only when needed to integrate over ghosts. A second problem arises due to the fact that we have chosen $\ddot{\gamma}$ as one of our dynamical variables. This is somewhat in contrast with the original idea of preserving the continuity properties of the path integral. We leave to future studies the analysis of the effect of this particular choice of dynamical variables. Third, we have considered the simplest possible scheme for taking the limit in which the higher order terms become less important in the action; with this choice the behaviors of the fourth and sixth order terms are locked together. A general approximation scheme in which the two terms may go to zero independently and introduce different corrections requires further study. Finally, note that we have only considered one specific sixth order term in the covariant action for gravity. Although a full calculation is needed, we do not expect the other sixth order terms to conspire and drastically change the results found here.

\section*{Acknowledgments}

This work was supported in part by NASA ATP grant NNX08AH27G, NSF grant PHY-0930521, and by Department of Energy grant DE-FG05-95ER40893-A020. MT is also supported by the Fay R. and Eugene L. Langberg chair.
 
\section*{Appendix}
\subsection{The Fourth Order Scalar System}
\label{4thorderappendix}

In the fourth order scalar case described by the Euclidean action
\be
S^E=\int d\tau \left( \frac{\alpha^2}{2} \phi''^2+\frac{1}{2}\phi'^2+\frac{m^2}{2}\phi^2 \right) \ ,
\ee
the wavefunctional is defined as 
\bea
\Psi_0[\phi_0,\phi_0'] &\equiv& \lim_{T\rightarrow \infty} \la (\phi_0,\phi_0';0)|(\phi_T,\phi_T';-T) \ra \nn \\ 
&=&N \exp \left[ -\frac{\sqrt{1-4m^2 \alpha^2}}{2(\lambda_2-\lambda_1)} \phi'^2 _0
+\frac{2m^2 \alpha - m}{\alpha (\lambda_2 -\lambda_1 )^2} \phi_0 \phi_0' 
-\frac{m \sqrt{1-4m^2 \alpha^2}}{2\alpha (\lambda_2-\lambda_1)}\phi_0^2 \right] \ ,
\eea
where $\lambda_1$ and $\lambda_2$ are found by solving the equations of motion

\be
\lambda_1 = \sqrt{\frac{1}{2\alpha^2} (1-\sqrt{1-4m^2 \alpha^2})} \ , \qquad
\lambda_2 = \sqrt{\frac{1}{2\alpha^2} (1+\sqrt{1-4m^2 \alpha^2})} \ .
\ee
The normalized probability, after integrating over $\phi'_0$ and rotating to Lorentzian time, becomes
\be
P[\phi_0]=\sqrt{ \frac{m}{\pi \alpha}\frac{\sqrt{1-4m^2 \alpha^2}}{(\lambda_2-\lambda_1)}} \exp\left[ -\frac{m}{ \alpha}\frac{\sqrt{1-4m^2 \alpha^2}}{(\lambda_2-\lambda_1)} \phi_0^2 \right] \ .
\ee

\subsection{Expansion of the 6th order action}

Starting from the action in equation (\ref{fullaction})
\bea
S&=&\int d^4 x \sqrt{-g} \mathcal{L} \ , \nn \\
\mathcal{L}&=&  M_{pl}^2  \left(\frac{R}{2}+\Lambda \right) +\lambda^2 R^2  -\alpha^2 C_{\mu \nu \rho \sigma}C^{\mu \nu \rho \sigma}
-  \frac{\beta^2}{M_{pl}^2} \nabla _\alpha R _{\mu\nu} \nabla ^\alpha R ^{\mu \nu} \ ,
\eea
the quadratic action for the fluctuations is found by varying the above twice. Since the first variation of the volume element is proportional to the trace of $\gamma$, and thus zero, the remaining terms can be written schematically as follows
\be
\delta^2 S = \int  d^4 x \left[ \delta^2 \sqrt{-g} \mathcal{L}+  \sqrt{-g} \left(
  \frac{M_{pl}^2}{2} \delta^2 R  +\lambda^2  \delta^2 R^2  -\alpha^2 \delta^2 C^2
-  \frac{\beta^2}{M_{pl}^2} \delta^2 \left(\nabla _\alpha R _{\mu\nu}\right) ^2 \right) \right] \ .
\ee
The first few terms, involving the Ricci Scalar, are standard, and the Weyl squared term was already calculated in~\cite{Clunan:2009er}, and can be written as
\be
\frac{1}{2} \delta^2 C_{\mu \nu \alpha \beta}C^{\mu \nu \alpha \beta}=
\frac{1}{2} e^{-4\rho}\left(\ddot{\gamma}_{ij}\ddot{\gamma}_{ij}
+2\ddot{\gamma}_{ij}\gamma_{ij,nn}+4\dot{\gamma}_{ij}\dot{\gamma}_{ij,nn}
+\gamma_{ij,nn}\gamma_{ij,mm} \right) \ .
\ee
Finally, the variation of the remaining sixth order term, with the aid of
\bea
\delta \left( \nabla_\alpha R_{\mu \nu} \right) &=& \half \delta_\alpha^0 \delta_\mu^i \delta_\nu^j \left[ \gamma_{ij}^{III}+2\dot{\gamma}_{ij}\left( \ddot{\rho}-2\dot{\rho}^2\right)-\dot{\gamma}_{ij,kk}
+2\gamma_{ij}\left(\rho^{III}+2\dot{\rho}\ddot{\rho}-4\dot{\rho}^3\right)
+2\dot{\rho}\gamma_{ij,kk} \right] \nn \\
&&+\half \delta_\alpha^k\left[ \left( \delta_\mu^0\delta_\nu^i+\delta_\mu^i\delta_\nu^0\right) 
\left( -\dot{\rho}\ddot{\gamma}_{ik}+2\dot{\gamma}_{ik}\left(\ddot{\rho}-2\dot{\rho}^2 \right)
+\dot{\rho} \gamma_{ik,jj}-4\gamma_{ik}
\left(\dot{\rho}^3-\dot{\rho}\ddot{\rho}\right)\right) \right. \nn \\
&&\left. +\delta_\mu^i \delta_\nu^j \left(  \ddot{\gamma}_{ij,k}+2\dot{\rho}\dot{\gamma}_{ij,k} -\gamma_{ij,kll}  \right)\right] \ ,
\eea
 gives
\bea
\delta^2 (\nabla_\alpha R_{\mu \nu}\nabla^\alpha R^{\mu \nu}) &=&   \half e^{-6\rho} \big[ -(\gamma^{III})^2+\ddot{\gamma}^2 (-10\dot{\rho}^2+4\ddot{\rho})
+3 \ddot{\gamma}^2_{,i} -3\dot{\gamma}^2_{,ij}+2\dot{\gamma}^2_{,i}(\ddot{\rho}+2\dot{\rho}^2) \nn \\
&+&\dot{\gamma}^2(4\rho^{IV}-4\dot{\rho}\rho^{III}-20\ddot{\rho}^2
+44\dot{\rho}^2\ddot{\rho}-48\dot{\rho}^4)
+\gamma_{,ijl}^2 +6\gamma_{,ij}^2(\dot{\rho}^2-\ddot{\rho}) \nn \\ &+&\gamma_{,i}^2 (-4\rho^{IV} -24\dot{\rho}^4 -8\ddot{\rho}^2 +72\dot{\rho}^2 \ddot{\rho} -4\dot{\rho} \rho^{III} ) \nn \\
 &+&\gamma^2(-8\rho^{VI}+336\dot{\rho}^4\ddot{\rho}+64\rho^{IV}\ddot{\rho}
+56\ddot{\rho}^3
-48\dot{\rho}^2\rho^{IV}-688\dot{\rho}^2\ddot{\rho}^2-304\dot{\rho}^3 \nn \\
&& \rho^{III}+36(\rho^{III})^2 
+56\dot{\rho}\rho^{V}+104\dot{\rho}\ddot{\rho}\rho^{III}+72\dot{\rho}^6)\big] \ .
\eea
Combining all of the above, we obtain the full expansion for the action, equation~(\ref{action6}).

\subsection{The Wavefunctional and Probability in a Minkowski Background}
 
The explicit form for the coefficients appearing in the definition of the wavefunctional, equation~(\ref{PsiM6}), can be cast in terms of 
$\lambda_i$ as follows
\bea
A_{00}&=&  \lambda _1 \lambda _2 \lambda _3 \left(-2 M_{pl}^2 \alpha ^2+\beta ^2 \left(-3 k^2+\lambda _1^2+\lambda _2^2+\lambda _3^2+\lambda _2 \lambda _3+\lambda _1 \lambda _2+\lambda _1\lambda _3\right)\right) \nn \\
A_{01}&=& M_{pl}^4+2 \alpha ^2 M_{pl}^2 \left(2 k^2+\lambda _2 \lambda _3+\lambda _1 \lambda _2+\lambda _1 \lambda _3\right) +\beta ^2 \left(3 k^4-\lambda _2^3 \lambda _3-\lambda _2^2 \lambda _3^2-\lambda _1^3 \left(\lambda _2+\lambda _3\right) \right. \nn \\
&& \left. -\lambda _1^2 \left(\lambda _2+\lambda _3\right){}^2-\lambda _1 \left(\lambda _2+\lambda _3\right) \left(-3 k^2+\lambda _2^2+\lambda _2 \lambda _3+\lambda _3^2\right)+\lambda _2 \left(3 k^2 \lambda _3-\lambda _3^3\right)\right)\nn \\
A_{02}&=& -2 \alpha ^2 M_{pl}^2 \left(\lambda _1+\lambda _2+\lambda _3\right)+\beta ^2 \left(\lambda _1^3+\lambda _2^3+\lambda _2^2 \lambda _3+\lambda _2 \lambda _3^2+\lambda _3^3+\lambda _1^2 \left(\lambda _2+\lambda _3\right) \right. \nn \\
&& \left. -3 k^2 \left(\lambda _1+\lambda _2+\lambda _3\right)+\lambda _1 \left(\lambda _2^2+\lambda _2 \lambda _3+\lambda _3^2\right)\right)\nn \\
A_{10}&=& -\beta ^2 \lambda _1 \lambda _2 \lambda _3 \left(\lambda _1+\lambda _2+\lambda _3\right) \nn \\
A_{11}&=&  \beta ^2 \left(\lambda _1+\lambda _2\right) \left(\lambda _1+\lambda _3\right) \left(\lambda _2+\lambda _3\right) \nn \\
A_{12}&=& 2 M^2_{pl} \alpha ^2-\beta ^2 \left(-3 k^2+\lambda _1^2+\lambda _2^2+\lambda _3^2+\lambda _1 \lambda _2+\lambda _1 \lambda _3+\lambda _2 \lambda _3\right) \nn \\
A_{20}&=& \beta ^2 \lambda _1 \lambda _2 \lambda _3 \nn \\
A_{21}&=& -\beta ^2 \left(\lambda _2 \lambda _3+\lambda _1 \lambda _2+\lambda _1 \lambda _3\right) \nn \\
A_{22}&=& \beta ^2 \left(\lambda _1+\lambda _2+\lambda _3\right) \ . 
\eea
The traced probability in Lorentzian time then reads
\bea
P[\gamma_0]&=&N N^* \exp \Bigg[
\frac{1}{4 M_{pl}^2 \beta ^2 (\lambda_1+\lambda_2+\lambda_3)}\left(-4 \beta ^2 \lambda_1 \lambda_2 \lambda_3 (\lambda_1+\lambda_2+\lambda_3) \left(-2 M_{pl}^2 \alpha ^2  \right.  \right. \nn \\
&& \left.+\beta ^2\left(-3 k^2+\lambda_1^2+\lambda_2^2+\lambda_2 \lambda_3+\lambda_3^2+\lambda_1 (\lambda_2+\lambda_3)\right)\right)+\left(-2 M^2 \alpha ^2 (\lambda_1+\lambda_2+\lambda_3) \right. \nn \\
&& +\beta ^2 \left(\lambda_1^3+\lambda_2^3+\lambda_3^3+\lambda_2^2 \lambda_3+\lambda_2 \lambda_3^2+\lambda_1^2 \lambda_2+\lambda_1^2\lambda_3+\lambda_1 (\lambda_2+\lambda_3)^2 \right. \nn \\
&& \left. \left. \left.-3 k^2 (\lambda_1+\lambda_2+\lambda_3)\right)\right)^2\right) |\gamma_0|^2 \Bigg]
\eea
\subsection{Classical Solutions and Probability in a de~Sitter Background}

The equations of motion in a de~Sitter background, equation~(\ref{dSEOM}), admit solutions in terms of Bessel functions. To find the general solution shown in the text, equation (\ref{dSsol}), it is convenient to search for a factorization of the full sixth order differential operator $D_6$ defined in equation~(\ref{dSD6}). $D_6$ can be split into a fourth order operator acting on a second order operator via
\be
D_6[z] \gamma_{cl}(z) = D_4[z] D_2[z] \gamma_{cl}(z) \ ,
\ee
where
\bea
D_4[z]&=& \frac{1}{z^2}\frac{d^4}{dz^4} - \left( \frac{2}{z^2} +\frac{1}{4z^4} \left( 25-4 \lambda_i^2
-8\frac{M_{pl}^2}{H^2}\frac{\alpha^2}{\beta^2}\right)\right)\frac{d^2}{dz^2}  +\frac{1}{2z^5}\left( 25-4 \lambda_i^2
-8\frac{M_{pl}^2}{H^2}\frac{\alpha^2}{\beta^2}\right) \frac{d}{dz}\nn \\
&&
+\frac{1}{z^2}-2\frac{M_{pl}^2}{H^2}\frac{\alpha^2}{\beta^2 z^4} +\frac{25-4\lambda_i^2}{z^4} \frac{1}{16 z^6 \beta^2 H^4}\left[ 8M_{pl}^4 - 8H^2 M_{pl}^2\left(\alpha^2-48\Lambda^2 - 4\alpha^2 \lambda_i^2 \right) \right. \nn \\
&& \left. +\beta^2 H^4 \left(153-104 \lambda_i^2 +16 \lambda^4 \right) \right]  \ ,\\
D_2[z]&=&z^2 \frac{d^2}{dz^2}-2z\frac{d}{dz}-\left(z^2+\lambda_i^2-\frac{9}{4}\right) \ .
\eea
Here, to avoid confusion, we have replaced the coefficient $\lambda^2$ of the $R^2$ term in the action with $\Lambda^2$, while $\lambda_i$ is a parameter in the decomposition. There are then three independent choices of the parameter $\lambda_i$, namely
\bea
\lambda_1&=&\frac{3}{2} \ , \\
\lambda_2&=& \frac{1}{2} \sqrt{\frac{13 H^2 \beta ^2+M_{pl}^2 \left(-4 \alpha ^2-2 \sqrt{\frac{4 H^4 \beta ^4+M_{pl}^4 \left(4 \alpha ^4-2 \beta ^2\right)-24 H^2 M_{pl}^2 \beta ^2 \left(\alpha ^2+4 \lambda ^2\right)}{M_{pl}^4}}\right)}{H^2 \beta ^2}} \ , \\
\lambda_3&=&\frac{1}{2} \sqrt{\frac{13 H^2 \beta ^2+M_{pl}^2 \left(-4 \alpha ^2+2 \sqrt{\frac{4 H^4 \beta ^4+M_{pl}^4 \left(4 \alpha ^4-2 \beta ^2\right)-24 H^2 M_{pl}^2 \beta ^2 \left(\alpha ^2+4 \lambda ^2\right)}{M_{pl}^4}}\right)}{H^2 \beta ^2}} \ ,
\eea
with these choices we obtain the six solutions of (\ref{dSsol}).

Once the classical solution is given, it is possible to calculate the associated value of the Euclidean action, find a wavefunctional as discussed in the text, and after tracing over the unobserved $\gamma''_0$ and $\gamma'_0$, and rotating back to Lorentzian time, eventually find a probability for $\gamma_0$. 

The full form of the non-normalized probability is then
\bea
&&P[\gamma_0]=N \exp \Big[ 
\left(k^2 \alpha ^3 M_{\text{pl}}^6 \left(\frac{3 H^4 k \beta ^4 \eta  r_2^5 \left(H k^3 \beta  \eta ^3+\alpha  \left(1+k^2 \eta ^2\right) M_{\text{pl}} r_2\right)}{M_{\text{pl}}^5}\right.\right. \nn \\
&&-\alpha ^3 \left(1+k^2 \eta ^2\right) r_1^8 \left(-\frac{H^2 k^3 \beta ^2 \eta ^3}{M_{\text{pl}}^2}+\frac{H \alpha  \beta  \left(2+k^2 \eta ^2+k^4 \eta ^4\right) r_2}{M_{\text{pl}}}+k \alpha ^2 \eta  \left(1+k^2 \eta ^2\right) r_2^2\right) \nn \\
&&+\frac{H^3 \beta ^3 r_1 r_2^4 \left(H^2 \beta ^2 \left(18+9 k^2 \eta ^2+4 k^4 \eta ^4\right)+H k \alpha  \beta  \eta  \left(6+7 k^2 \eta ^2+k^6 \eta ^6\right) M_{\text{pl}} r_2 \right)}{M_{\text{pl}}^5} \nn \\
&&+\frac{H^3 \beta ^3 r_1 r_2^4 \left(\alpha ^2 \left(1+k^2 \eta ^2+k^4 \eta ^4+k^6 \eta ^6\right) M_{\text{pl}}^2 r_2^2\right)}{M_{\text{pl}}^5} \nn \\
&&-\alpha ^2 r_1^7 \Bigg(-\frac{H^3 k^6 \beta ^3 \eta ^6}{M_{\text{pl}}^3}-\frac{H^2 k^3 \alpha  \beta ^2 \eta ^3 \left(12+17 k^2 \eta ^2+3 k^4 \eta ^4\right) r_2}{M_{\text{pl}}^2} \nn \\
&&+\frac{H \alpha ^2 \beta  \left(8+20 k^2 \eta ^2+21 k^4 \eta ^4+9 k^6 \eta ^6\right) r_2^2}{M_{\text{pl}}}+k \alpha ^3 \eta  \left(1+k^2 \eta ^2\right)^2 r_2^3\Bigg) \nn \\
&&+\frac{1}{M_{\text{pl}}^5}H^2 \beta ^2 r_1^2 r_2^3 \left(H^3 \beta ^3 \left(18+9 k^2 \eta ^2+k^4 \eta ^4\right)+H^2 k \alpha  \beta ^2 \eta  \left(18+29 k^2 \eta ^2+10 k^4 \eta ^4+k^6 \eta ^6\right) M_{\text{pl}} r_2\right. \nn \\
&&\left.+H \alpha ^2 \beta  \left(1+k^2 \eta ^2+5 k^4 \eta ^4+2 k^6 \eta ^6\right) M_{\text{pl}}^2 r_2^2+k \alpha ^3 \eta  \left(4+5 k^2 \eta ^2+k^4 \eta ^4\right) M_{\text{pl}}^3 r_2^3\right) \nn \\
&&-\frac{1}{M_{\text{pl}}^4}H \alpha  \beta  r_1^6 \left(3 H^3 k \beta ^3 \eta  \left(1+k^2 \eta ^2\right)+H^2 \alpha  \beta ^2 \left(1+k^2 \eta ^2+k^4 \eta ^4-12 k^6 \eta ^6-4 k^8 \eta ^8\right) M_{\text{pl}} r_2\right. \nn \\
&&\left.-H k \alpha ^2 \beta  \eta  \left(-1+17 k^2 \eta ^2+21 k^4 \eta ^4+k^6 \eta ^6\right) M_{\text{pl}}^2 r_2^2+\alpha ^3 \left(6+17 k^2 \eta ^2+18 k^4 \eta ^4+7 k^6 \eta ^6\right) M_{\text{pl}}^3 r_2^3\right) \nn \\
&&+\frac{1}{M_{\text{pl}}^5}H \beta  r_1^3 r_2^2 \left(H^4 \beta ^4 \left(18+45 k^2 \eta ^2+19 k^4 \eta ^4\right)+H^2 k^4 \alpha ^2 \beta ^2 \eta ^4 \left(21+9 k^2 \eta ^2+k^4 \eta ^4\right) M_{\text{pl}}^2 r_2^2\right. \nn \\
&&\left.+2 H k \alpha ^3 \beta  \eta  \left(12+17 k^2 \eta ^2+5 k^4 \eta ^4+k^6 \eta ^6\right) M_{\text{pl}}^3 r_2^3+\alpha ^4 \left(1+k^2 \eta ^2+k^4 \eta ^4+k^6 \eta ^6\right) M_{\text{pl}}^4 r_2^4\right) \nn \\
&&+r_1^5 \left(\frac{H^5 \beta ^5 \left(6+4 k^2 \eta ^2-3 k^4 \eta ^4\right)}{M_{\text{pl}}^5}-\frac{H^4 k \alpha  \beta ^4 \eta  \left(3+2 k^2 \eta ^2-2 k^4 \eta ^4+k^6 \eta ^6\right) r_2}{M_{\text{pl}}^4}\right. \nn \\
&&+\frac{H^3 \alpha ^2 \beta ^3 \left(-1-k^2 \eta ^2-2 k^4 \eta ^4+10 k^6 \eta ^6+4 k^8 \eta ^8\right) r_2^2}{M_{\text{pl}}^3} \nn \\
&&+\frac{H^2 k \alpha ^3 \beta ^2 \eta  \left(18+39 k^2 \eta ^2+26 k^4 \eta ^4+5 k^6 \eta ^6\right) r_2^3}{M_{\text{pl}}^2}+\frac{2 H k^4 \alpha ^4 \beta  \eta ^4 \left(1+k^2 \eta ^2\right) r_2^4}{M_{\text{pl}}} \nn \\
&&+k \alpha ^5 \eta  \left(1+k^2 \eta ^2\right)^2 r_2^5\Bigg) \nn \\
&&+\frac{1}{M_{\text{pl}}^5}r_1^4 r_2 \left(H^5 \beta ^5 \left(24+49 k^2 \eta ^2+16 k^4 \eta ^4\right)+H^4 k^3 \alpha  \beta ^4 \eta ^3 \left(16+27 k^2 \eta ^2+9 k^4 \eta ^4\right) M_{\text{pl}} r_2\right. \nn \\
&&+H^3 k^4 \alpha ^2 \beta ^3 \eta ^4 \left(18+9 k^2 \eta ^2+k^4 \eta ^4\right) M_{\text{pl}}^2 r_2^2+H^2 k \alpha ^3 \beta ^2 \eta  \left(39+57 k^2 \eta ^2+20 k^4 \eta ^4+4 k^6 \eta ^6\right) M_{\text{pl}}^3 r_2^3 \nn \\
&&\left.\left.+H \alpha ^4 \beta  \left(1+k^2 \eta ^2+4 k^4 \eta ^4+4 k^6 \eta ^6\right) M_{\text{pl}}^4 r_2^4+k \alpha ^5 \eta  \left(1+k^2 \eta ^2\right)^2 M_{\text{pl}}^5 r_2^5\right)\Bigg)\Bigg)\right/ \nn \\
&&\left(H^2 \beta ^2 \eta  \left(1+k^2 \eta ^2\right) \left(r_1-r_2\right) \left(-H^4 \beta ^4 \left(3+2 k^2 \eta ^2\right) r_1-H^4 \beta ^4 \left(3+2 k^2 \eta ^2\right) r_2\right.\right. \nn \\
&&+\alpha ^3 \left(1+k^2 \eta ^2\right) M_{\text{pl}}^3 r_1^4 \left(H k^3 \beta  \eta ^3+\alpha  \left(1+k^2 \eta ^2\right) M_{\text{pl}} r_2\right) \nn \\
&&+H k^3 \alpha ^2 \beta  \eta ^3 M_{\text{pl}}^2 r_1^2 r_2 \left(H k^3 \beta  \eta ^3+\alpha  \left(1+k^2 \eta ^2\right) M_{\text{pl}} r_2\right)+\alpha ^2 M_{\text{pl}}^2 r_1^3 \left(H^2 k^6 \beta ^2 \eta ^6\right. \nn \\
&&\left.\left.\left.+3 H k^3 \alpha  \beta  \eta ^3 \left(1+k^2 \eta ^2\right) M_{\text{pl}} r_2+\alpha ^2 \left(1+k^2 \eta ^2\right)^2 M_{\text{pl}}^2 r_2^2\right)\right)\right) \ ,
\eea
where $r_1$ and $r_2$ are given by
\be
r_{1,2}=\sqrt{1\pm \sqrt{1-\frac{\beta ^2}{2\alpha ^4}}} \ .
\ee


\end{document}